\documentclass[lettersize,journal]{IEEEtran}
\usepackage{amsmath,amsfonts}
\usepackage{algorithmic}
\usepackage{algorithm}
\usepackage{array}
\usepackage[caption=false,font=normalsize,labelfont=sf,textfont=sf]{subfig}
\usepackage{textcomp}
\usepackage{stfloats}
\usepackage{url}
\usepackage{verbatim}
\usepackage{graphicx}
\usepackage{cite}
\usepackage{xcolor}
\usepackage{soul}
\begin{document}

\title{Experimental Evaluation of All-Optical Up- and Down-Conversion of 3GPP 5G NR Signals using an Optomechanical Crystal Cavity Frequency Comb}

\author{Vicente Fito*, Raúl Ortiz*, Maria Morant,~\IEEEmembership{Member,~IEEE,} Laura Mercadé, Roberto Llorente,~\IEEEmembership{Member,~IEEE,} and Alejandro Martínez,~\IEEEmembership{Senior Member,~IEEE}

\thanks{Manuscript received December 18, 2023. Reviewed February 14, April 16 and May 29, 2024.}
\thanks{This work was supported in part by Generalitat Valenciana CIAICO/2021/201 TERAFLEX project, IDIFEDER/2021/061 and Spanish Ministry of Science and Innovation under grants ALLEGRO (PID2021-124618NB-C21) and MUSICIAN (CHISTERA IV Cofund 2021) and the European Commission under grant MAGNIFIC (HORIZON-CL4-2022-RESILIENCE-01-10 101091968). *V. Fito and R. Ortiz contributed equally to this work, supported by PAID-01-20 UPV and GVA CIACIF/2021/006 grants, respectively. M. Morant work supported by Spain R\&D project PID2019-106163RJ-I00/AEI/10.13039/501100011033 MULTICORE+.}
\thanks{Authors are with Valencia Nanophotonics Technology Center,  Universitat Polit\`{e}cnica de Val\`{e}ncia, 46022 Valencia, Spain. (\mbox{e-mail}: \{vfitest, rortfer, mmorant, laumermo, rllorent, amartinez\}@ntc.upv.es). }}

\markboth{Journal of Lightwave Technology}%
{Fito \MakeLowercase{\textit{et al.}}: Title}

\maketitle

\begin{abstract}
Optomechanical crystal cavities (OMCCs) allow the interaction between localized optical and mechanical modes through the radiation-pressure force. Driving such cavities with blue-detuned lasers relative to the optical resonance can induce a phonon lasing regime where the OMCC supports self-sustained mechanical oscillations. This dynamic state results in a narrow and stable microwave tone that modulates the laser at integer multiples of the mechanical resonance frequency, ultimately creating an optomechanical (OM) frequency comb suitable for microwave photonics applications. OMCCs enable compact, low-cost power-efficient all-photonic processing of multiple microwave signals, crucial for current 5G and future beyond-5G systems, whilst being compatible with silicon integrated photonic circuits. This work reports the experimental demonstration of all-optical multi-frequency up- and down-conversion of 3GPP 5G new-radio (NR) signals from the low- to mid- and extended-mid bands using the first and second harmonics of the frequency comb generated in a silicon OMCC. The OM comb generates up to 6 harmonics in the K-band, which is suitable for microwave photonic applications. The experimental demonstration also evaluates the impact of the phase-noise and the signal-to-noise ratio (SNR) in the frequency-converted 5G NR signals when the first and second OMCC harmonics are employed for frequency conversion.  
\end{abstract}


\section{Introduction}
\IEEEPARstart{C}{avity} optomechanics explore the interaction between light and mechanical waves when confined in small cavities, leading to a range of interesting physical phenomena \cite{CavOpt, Chan, Verhagen}. When an optomechanical crystal cavity (OMCC) is driven by a laser whose frequency is larger than the optical resonance (i.e. blue-detuned), the cavity can enter into a dynamics in which the optical mode interacts with the mechanical motion reducing its damping rate, which ultimately leads to a regime of self-sustained oscillations which effectively generates mechanical phonon-based lasing \cite{KIP08-SCI, Navarro-Urrios}. \\
\indent 
The OMCC operates in a highly non-linear regime where a very narrow tone at the mechanical resonance frequency modulates the driving laser \cite{Lasing1,Lasing2}. Interestingly, this new optical signal is also modulated by the amplified mechanical motion \cite{CavOpt}, resulting in a cascaded process that resembles the formation of an optical frequency comb \cite{Microwave, Miri}. Ad-hoc design of the cavity permits obtaining a mechanical resonance at GHz frequencies suitable for microwave photonic applications, where a set of very narrow and highly stable microwave tones (at integer multiples of the mechanical frequency) are generated after photodetection. \\
\indent 
Frequency conversion using photonic techniques have been recently introduced due to their potential advantages to provide large  bandwidth and full-band operation enabling the operation at extended radio-frequency (RF) bands.  Optoelectronic oscillator (OEO) systems have been also proposed for frequency conversion, as they enable a power-efficient frequency conversion \cite{CAP07-NP}. Silicon photonic integrated OEOs have been demonstrated with phase noise levels of $-80$\;dBc/Hz at 10 and 100\;kHz frequency offsets \cite{LowPhaseNoise}. Also microwave photonic mixers can implement frequency conversion using two I/Q modulators as proposed in \cite{IQModUp}. Although photonic solutions using modulators have a larger footprint, the phase noise level can be reduced to $-90$\;dBc/Hz using two cascaded electro-absorption modulators (EAMs) \cite{2EAMs}. In this work, we also evaluate the phase noise of the first and second harmonics generated with the proposed OMCC, achieving levels smaller than $-97$\;dBc/Hz and $-91$\;dBc/Hz at 100\;kHz for the first and second harmonics, respectively, while being more a compact solution than other implementations reported in the literature \cite{HOS08-IEEE}. The compactness of the solution is of special importance for large antenna arrays employed in massive MIMO wireless communication systems, where the integration of numerous antenna elements demands compact, low driving power and efficient processing chains. As large antenna arrays are becoming a reality in 5G and beyond-5G radio networks \cite{BeamformingArray}, the associated up/down-conversion stages in the processing chain remain a challenging technology where bulky and expensive electronic solutions are often employed \cite{MIMO}. For these reasons, the herein proposed OMCC offers great potential for microwave photonic applications, as it enables the optical frequency conversion of antenna signals with a compact, power-efficient and reduced phase-noise device that can be further integrated with other photonic and electronic systems employing silicon photonics micro- and nano-fabrication processes. \\
\begin{figure*}[!b]
  \centering
    \includegraphics[width=18.3cm]{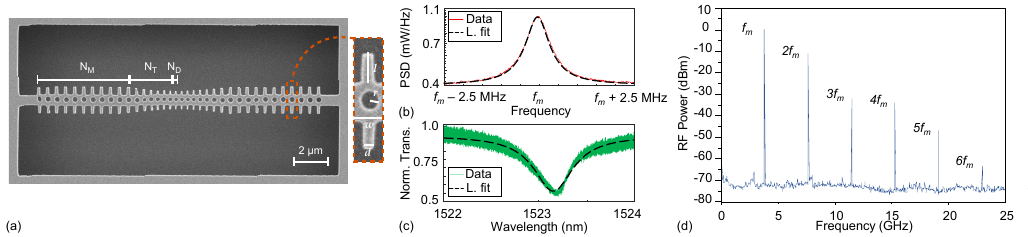}
     \vspace{-0.7cm}
\caption{\fontsize{8}{10}\selectfont\centering (a)\;SEM image of the silicon OMCC composed by 10 mirror cells ($N_M$), 6 transition cells ($N_T$) and a defect cell ($N_D$). Inset: Cavity unit cell with its dimensions. (b)\;PSD of the thermally transduced mechanical mode with a Lorentzian fit. (c)\;Optical mode ($\lambda_o$) measured with an oscilloscope and a Lorentzian fit curve. (d)\;RF spectrum with six harmonics of the OM frequency comb in the lasing regime (measured with RBW=10\;MHz).}
\label{fig:estabilidad}
\end{figure*}
OMCCs have also been proposed for microwave-to-optical conversion of single photons as quantum transducers \cite{Transd1, Transd2}. This could facilitate the transfer of quantum states between distant superconducting quantum processors. However, OMCCs have a role beyond quantum transduction \cite{Verhagen} as they can be used for frequency conversion of telecommunication signals \cite{HOS08-IEEE, LyM}. In previous works, we demonstrated the frequency conversion using the first harmonic generated by a silicon OMCC considering orthogonal frequency division multiplexing (OFDM) signals following the IEEE 802.16e WiMAX wireless standard with up to \mbox{7-MHz} bandwidth (BW) \cite{LyM} and for SATCOM applications \cite{LyE}. In this work, we extend the experimental evaluation of all-optical multi-frequency conversion using the first and the second harmonics via cavity optomechanics in the phonon lasing regime with higher-BW wireless signals following the $3^{rd}$ Generation Partnership Project (3GPP) 5G New-Radio (NR) specification with up to 100-MHz BW per channel. The novelty of this work resides in using the on-chip silicon OMCC to perform frequency up- and down-conversion within the low ($<1$\;GHz), mid (1-6\;GHz) and extended mid-bands ($8.5$\;GHz) of the 5G spectrum. The 7 to 8.5\;GHz spectrum range was recently (in December 2023) proposed by the ITU for next-generation 6G to provide immersive, reliable, and low-latency communication (e.g., industrial applications, telemedicine...) as well as enhanced ubiquitous connectivity to bridge the digital divide in rural and remote areas. We evaluate all the 3GPP 5G standard channel BWs ranging from 5 to 100\;MHz and analyze experimentally the impact of the OMCC frequency offset and the resulting signal-to-noise ratio (SNR) levels of the frequency-converted 5G NR signals, comparing the received error vector magnitude (EVM) with the 3GPP recommendation.\\ 
\indent 
This paper is structured as follows: In Section II, the designed and fabricated silicon OMCC is presented and characterized experimentally including the short- and long-term evaluation of the mechanical frequency offset for the first and second harmonics of the OMCC, together with phase noise measurements to identify the possible noise sources in the oscillator. In Section III, the performance of the all-optical frequency conversion of 5G NR signals using the first and second OMCC harmonics is evaluated in both up- and down-conversion processes, taking into account the oscillator frequency offset and the SNR level of the frequency-converted 5G NR signals. Finally, in Section IV, the main conclusions of this work are highlighted. Our results open the door toward the use of OMCCs integrated in silicon chips as appropriate building blocks for 5G and beyond-5G distribution networks.

\section{Silicon OMCC Design and Characterization}
The cavity used to conduct the experimental demonstration is a suspended silicon OMCC \cite{Microwave}. The silicon OMCC is fabricated using standard silicon fabrication tools \cite{Jordi}. Its architecture features periodic photonic crystal mirror cells along its sides, establishing a photonic bandgap for TE-like modes at $\lambda_o=1523.2$\;nm. The OMCC structure composed by 10 mirror cells ($N_M$), 6 transition cells ($N_T$), and a defect cell ($N_D$) is shown in the scanning electron microscope (SEM) image included in Fig.\;\ref{fig:estabilidad}(a) within a foot-print of $\approx 10\;\mu m^{2}$. A complete phononic bandgap is achieved at the mirror cells at $f_m=3.8$\;GHz \cite{Microwave}. The mirror cell sizes are: $l_M=500$\;nm, $d_M=250$\;nm, $r_M=150$\;nm, $a_M=500$\;nm. To increase the interaction between the optical and mechanical modes, the unit cell dimensions adiabatically vary from the cavity sides to its center so that the optical and mechanical modes are confined at the center of the cavity. The defect cell size is $l_D=250$\;nm, $d_D=163$\;nm, $r_D=98$\;nm, $a_D=325$\;nm. The waveguide width and thickness are $w=570$\;nm and $t=220$\;nm.\\
\indent 
It is possible to calculate this interaction with the vacuum OM coupling rate ($g_0$), which has two contributions: the photoelastic (PE) and the moving interfaces (MI) effects \cite{gOM}.
Our cavity supports a set of GHz mechanical resonances \cite{Engineering}, each of them having a different vacuum OM coupling rate with the optical mode. One of them will show the highest value of $g_0$ ($g_0\approx2\pi\cdot200$\;kHz, as reported in \cite{Microwave}) and this will be the mode used in this experimental demonstration. 

\begin{figure*}[t]
  \centering
    \includegraphics[height=5.5cm]{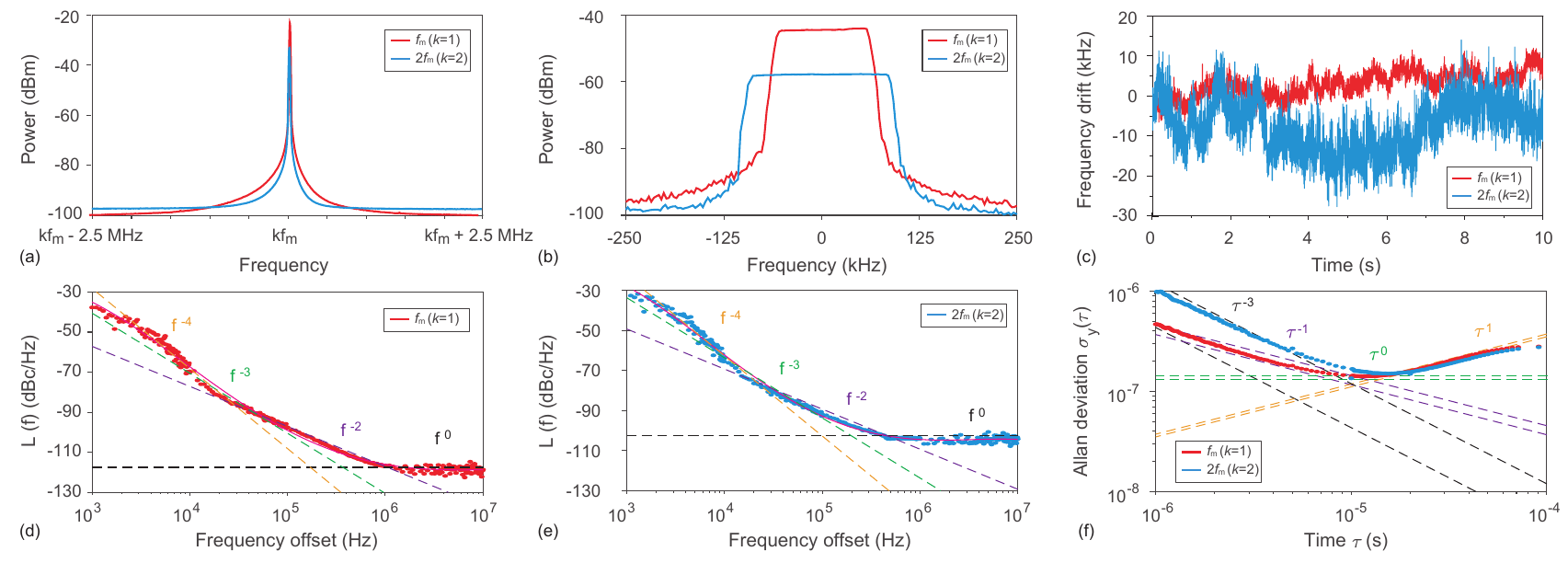}
     \vspace{-0.4cm}
\caption{\fontsize{8}{10}\selectfont\centering (a)\;Measured RF power of the first two harmonics of the OM frequency comb generated with the OMCC in the lasing regime at $k f_m$. (b)\;Mechanical frequency drift measured with max hold during 20 minutes (long-term evaluation) for the first and second harmonics. (c)\;Frequency offset of the first and second OMCC harmonic measured over 10\;s (short-term evaluation). (d, e)\;Phase noise measurements for the first and second harmonics of the GHz lasing regime for $k=1$ and 2. The fitted slopes correspond to the noise sources in the oscillator. (f)\;Allan deviation results calculated from the phase noise contributions for the first and second harmonics. Blue and red continuous lines show the Allan deviation, and dashed lines depict their power low fits.}
\label{fig:drifts}
\end{figure*}

\indent In order to characterize the optical and the mechanical response of the cavity, light is coupled in and out of the system by means of a tapered fiber \cite{Hill} via a fiber loop \cite{Jordi}. The resulting optical resonance was observed in transmission on an oscilloscope, and the mechanical response was evaluated through reflection using a circulator and monitored using a spectrum analyzer. Fig.\;\ref{fig:estabilidad}(b)-(c) show the thermally transduced mechanical mode at $f_{m}=3.84$\;GHz and the normalized transmission of the measured optical resonance of the cavity centered at $\lambda_o=1523$\;nm. The mechanical and optical quality factors are: $Q_m=6.7\cdot10^3$ and $Q_o=3.0\cdot10^3$. This leads to an optical decay rate $\kappa=65$\;GHz, which means that we are in the unresolved sideband regime ($\kappa >> 2\pi f_{m}$). In this given cavity limit, an optomechanical frequency comb can arise \cite{Miri} as it can be seen experimentally in Fig.\;\ref{fig:estabilidad}(d), showing the first six harmonics of the OM frequency comb in the so-called phonon lasing regime spaced by the mechanical frequency at $kf_m$ (with $k=1, 2, ...6$). With the harmonics provided by the OM frequency comb, the microwave photonics RF spectrum up to the K-band can be covered. Next, a detailed examination of the Voigt-like shape characterizing the first and second harmonics in the lasing regime mode is presented in Fig.\;\ref{fig:drifts}(a). The maximum amplitude span and frequency drift of these mechanical peaks, measured over a twenty-minute interval (long-term evaluation), are illustrated in Fig.\;\ref{fig:drifts}(b). Notably, the frequency offset of the second harmonic ($k=2$) is approximately twice that of the first harmonic ($k=1$), yet both exhibit frequency offsets smaller than 250\;kHz. These values are well below the specification defined by 3GPP in TS 38.141-1 (Release 15)\cite{3GPP}, where the maximum frequency offset is set for both non-contiguous and in-band adjacent channel configurations. In particular, for non-contiguous channels, a maximum frequency offset of 2.5\;MHz is allowed for 5-MHz BW 5G signals. This frequency offset can increase up to 30\;MHz for 100-MHz BW 5G signals. The selectivity is more restrictive for in-band adjacent channels where a maximum frequency offset of ±2.5025\;MHz (i.e. a maximum of 5.005\;MHz) is allowed for 5-MHz BW 5G signals and up to ±9.4675\;MHz (i.e. maximum of 18.935\;MHz) for 100-MHz BW 5G signals \cite{3GPP}. \\
\indent In order to comprehensively evaluate the system's stability, we conducted measurements for phase noise and the temporal evolution of the center frequency at short periods. An examination of the system's temporal evolution of the detected frequency was conducted covering a time span of 10\;seconds (short-term evaluation). The results of this analysis are presented in Fig.\;\ref{fig:drifts}(c), where the frequency drift of both harmonics is assessed over the specified duration. Notably, the frequency deviation of $kf_m$ for the second harmonic is observed to be approximately twice that obtained for the first harmonic, aligning with the anticipated behavior of Fig.\;\ref{fig:drifts}(b) at longer time measurements. Figures\;\ref{fig:drifts}(d)-(e) depict the phase noise of the first and second harmonics, along with their dependencies based on Lesson's model \cite{Rubiola}. The phase noise exhibits characteristic dependencies, including $f^{-4}$ (random walk of frequency), $f^{-3}$ (flicker frequency noise), $f^{-2}$ (white frequency noise or random walk), and $f^{0}$ (white phase noise). The lower white phase noise observed in Fig.\;\ref{fig:drifts}(e) can be attributed to the lower peak power of the signal. Further analysis involves retrieving the mechanical linewidth of the harmonics from the white frequency noise dependence ($f^{-2}$), resulting in an effective linewidth of $\Gamma_{eff}/2\pi=74$\;Hz and lower values of the phase noise of -97.1\;dBc/Hz at 100\;kHz for the first harmonic. This finding emphasizes the system's highly competitive, stable free-running oscillator \cite{Remy}. The jitter of the system \cite{jitter} for the two given harmonics was also calculated providing values as low as $J_{RMS}=1.1$\;ps at 10\;kHz - 100\;kHz for both comb lines, resulting in a thirtyfold reduction in jitter compared to other free running oscillators \cite{jitter_comp}. The stability of the system is further explored through its representation in the Allan deviation included in Fig.\;\ref{fig:drifts}(f) for both comb peaks. A noteworthy observation is the convergence of the random walk of frequency ($\tau^{1} - f^{-4}$ contribution) for both harmonics. This convergence indicates that the random processes contributing to this phenomenon are comparable for both harmonics. This similarity in behavior can be attributed to various factors, such as thermal noise or other stochastic processes influencing the system \cite{Rubiola}.

\section{All-optical frequency conversion of 5G NR signals using the OM Frequency Comb}

\begin{figure}[t]
  \centering
    \includegraphics[width=7.5cm]{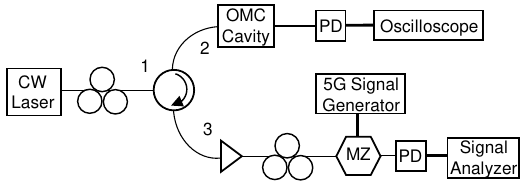}
     \vspace{-0.3cm}
\caption{\fontsize{8}{10}\selectfont\centering Experimental setup for all-optical frequency conversion using OMCC. The transmission path is used for phonon lasing monitoring, and the reflection path is used for 5G NR transmission implementing up- or down-conversion.}
\label{fig:setup}
\end{figure}

Figure\;\ref{fig:setup} shows the experimental setup used for the evaluation of all-optical frequency conversion performance of 5G NR signals using the proposed OMCC. The primary objective of this experimental demonstration is the frequency conversion of fully standard 3GPP 5G NR signals from the 5G low band (below 1\;GHz) --characterized by broader coverage yet reduced speeds-- to both the 5G mid-band (ranging from 1 to 6\;GHz) and the extended mid-band (up to 8.5\;GHz) also proposed by ITU for 6G. This is accomplished by leveraging the OM frequency comb generated by the OMCC when operating in the phonon lasing regime. When the continuous wave (CW) light interacts with the OMCC, transmitted and reflected components are obtained. The transmitted beam is used for monitoring the mechanical mode of the OMCC, while the reflected beam is amplified with an Erbium-doped fiber amplifier (EDFA) and modulated using a Mach-Zehnder (MZ). The MZ is fed with  standard 3GPP 5G NR signals using frequency division duplex (FDD) in the frequency range FR1, generated with a vector signal generator (R\&S SMW200A), centered at $f_{0}$ with electrical BWs up to 100\;MHz. Optical double sideband modulation is achieved with the MZ operating at the quadrature bias (QB) point, generating replicas of the original 5G signal at linear combinations of the OMCC harmonic frequencies (obtained at center $kf_{m}$ with $k=1$ and $k=2$) and the 5G signal original frequency ($f_{0}$). Both upconversion and downconversion frequency processes are evaluated considering the same frequency ranges for 5G low-, mid-, and extended mid-bands. Thus, the original 5G signal center frequency $f_{0}$ is changed accordingly. \\ 
\indent At the receiver, a 12-GHz BW photodetector (PD) is used, and the resulting up- and down-converted 5G NR signals are demodulated with R\&S FSW43 signal and spectrum analyzer. The error vector magnitude (EVM) and received constellations of the 5G physical downlink shared channel (PDSCH) are analyzed and compared with the 3GPP standard EVM recommendation for 5G PDSCH modulated with quadrature phase shift keying (QPSK) of 18.5\%\cite{3GPP}. 
\subsection{Frequency upconversion performance}
\begin{figure}[!b]
  \centering
    \includegraphics[width=8.5cm]{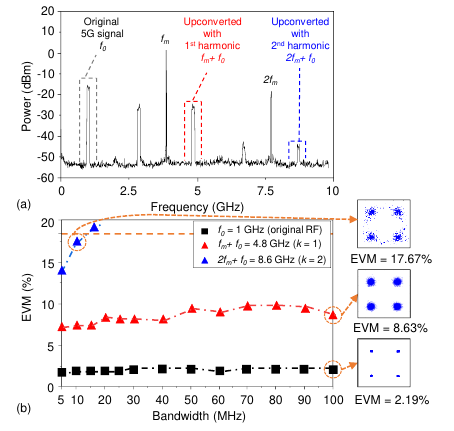}
     \vspace{-0.4cm}
\caption{\fontsize{8}{10}\selectfont\centering (a)\;Received RF spectrum of the upconversion process of a 100-MHz BW 5G NR signal originally at $f_{0}=1\;$GHz (measured with RBW = 10 MHz) and optical power received by the PD $= -2$\;dBm. (b)\;Measured EVM and constellations of the PDSCH upconverted signals at $k f_{m}+f_{0}$ using the first ($k=1$) and second ($k=2$) \mbox{OMCC harmonics} vs. BW, compared with the original 5G NR signal at $f_{0}=1$\;GHz. The 3GPP EVM recommendation is included as a horizontal dashed line.}
\label{fig:upconversion}
\end{figure}

The resulting RF spectrum for upconversion applications is represented in Fig.\;\ref{fig:upconversion}(a), showing the original low-band 5G NR signal centered at $f_{0}=1$\;GHz and the first OMCC harmonic ($k=1$) located at $f_{m}=3.84$\;GHz, with double-sideband replicas corresponding to the 5G mid-band centered at \mbox{$f_{m}-f_{0}=2.84$\;GHz} and \mbox{$f_{m}+f_{0}=4.84$\;GHz}. Next to this, the second OMCC harmonic ($k=2$) is found at \mbox{$2f_{m}=7.68$\;GHz}, and the upconverted signals appear at \mbox{$2f_{m}-f_{0}=6.68$\;GHz} and $2f_{m}+f_{0}=8.68$\;GHz (corresponding to the 5G extended mid-band). In these experiments, the worst-case scenario is considered by measuring the EVM of the higher sidebands located at $f_{m}+f_{0}=4.84$\;GHz (upconverted using the first harmonic with $k=1$) and at $2f_{m}+f_{0}=8.68$\;GHz (upconverted using the second harmonic with $k=2$).

Figure\;\ref{fig:upconversion}(b) presents the received EVM for each upconverted frequency and for different electrical bandwidths following the 3GPP 5G NR standard ranging from 5-MHz to 100-MHz BW and adjusting the electrical power of the generated signals to obtain the optimum EVM  (within the range of the 5G NR generator). The figure includes examples of received constellations of the 5G physical downlink shared channel (PDSCH). A horizontal dashed line in the EVM graphs represents the 3GPP standard EVM recommendation for 5G PDSCH modulated with QPSK at 18.5\%\cite{3GPP}. The experimental results confirm that the upconverted signals to the 5G mid-band using the first OMCC harmonic meet the 3GPP EVM recommendation for the maximum 5G BW of 100\;MHz. However, the upconversion to the extended mid-band using the second harmonic is limited to 5G NR signals with up to 10-MHz BW. The EVM degradation is more pronounced for the signals upconverted using the second harmonic due to its reduced SNR. As it can be observed in the resulting spectrum included in Fig.\;\ref{fig:upconversion}(a), the SNR decreases from $35.6$\;dB for the original 5G signal ($f_{0}$) to $26.6$\;dB for the upconverted signal with the first harmonic ($f_{m}+f_{0}$), further diminishing to $8.6$\;dB for the upconverted signal with the second harmonic ($2f_{m}+f_{0}$). This difference in SNR between the signal upconverted with the first and second harmonics, together with the higher frequency offset present in the second harmonic, as seen in Fig.\;\ref{fig:upconversion}(b), imposes limitations on upconversion using the second harmonic to 5G bandwidths up to 10\;MHz. Achieving higher bandwidths with the second harmonic will require a higher SNR, as evaluated in the next subsections of this study.

\subsection{Frequency downconversion performance}

\begin{figure}[t]
  \centering
\includegraphics[width=7.5cm]{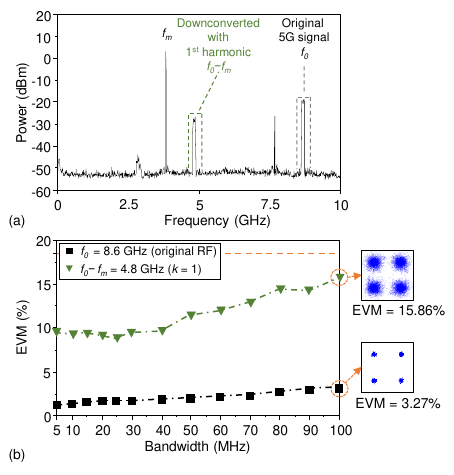}
     \vspace{-0.4cm}
\caption{\fontsize{8}{10}\selectfont\centering (a)\;Received RF spectrum of the downconversion process with original 100-MHz BW 5G NR signal centered at $f_0=8.6\;$GHz and downconverted to $4.8\;$GHz (measured with RBW=10 MHz) with optical power received at the PD$=-2\;$dBm. (b)\;Received EVM and example constellations of the PDSCH of the original 5G NR and downconverted signals using the first OMCC harmonic for different electrical BWs. \mbox{3GPP EVM} recommendation included as a horizontal dashed line.}
\label{fig:downconversion}
\end{figure}

The downconversion process is also evaluated using the same experimental setup depicted in Fig.\;\ref{fig:setup}. In order to compare the same frequency ranges, in this case, the original 5G NR signal is generated at $f_{0}=8.68\;$GHz. Thus, the downconverted signal using the first OMC harmonic appears at $f_{0}-f_{m}=4.84\;$GHz as shown in the received spectrum reported in Fig.\;\ref{fig:downconversion}(a). The results of Fig.\;\ref{fig:downconversion}(b) confirm the EVM compliance of 5G NR signals with up to 100-MHz BW downconverted employing the first harmonic.

If we compare the up- and down-conversion processes, we can observe that the received EVM escalates from 7.36\% for a 5-MHz BW to 8.63\% for a 100-MHz BW for signals upconverted with the first harmonic. For downconverted signals also using the first harmonic, the received EVM rises from 9.6\% for a 5-MHz BW to 15.86\% for a 100-MHz BW. It is worth noting that the SNR of the upconverted signal with the first harmonic is measured to be $26.6$\;dB, in contrast with the downconverted signal at the same center frequency, which records an SNR of $24.6$\;dB. Consequently, the EVM of the upconverted signals exhibits a marginally better performance than the downconverted signals over the same frequencies. The impact of the frequency offset and received SNR in the EVM will be further evaluated in the next subsections. 

\subsection{Impact of frequency offset}
\begin{figure}[t]
  \begin{flushleft}
\includegraphics[width=9cm]{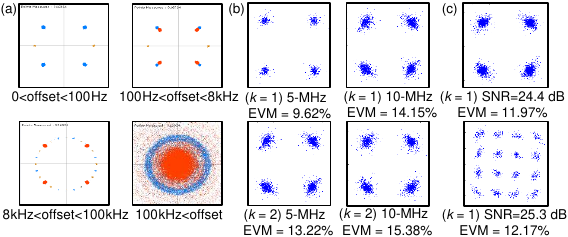}
     \vspace{-0.6cm}
\caption{\fontsize{8}{10}\selectfont\centering (a)\;Received 5G NR constellations when transmitted at $f_{0}=1\;$GHz and received at $f_{0}+f$\textsubscript{offset} to evaluate the analyzer sensitivity. (b)\;Received PDSCH constellations and EVM of the upconverted \mbox{5G NR} signals using the first harmonic ($k=1$ to $4.8\;$GHz) and the second harmonic ($k=2$ to $8.6\;$GHz) for different BWs and constant SNR\;=\;25\;dB. (c) Comparison of PDSCH QPSK and 16QAM up-converted 5G NR signals with  $k=1$.}
\label{fig:Constellations}
    \end{flushleft}
\end{figure}

In Fig.\;\ref{fig:Constellations}(a), we examine the impact of frequency offset on the received 5G NR constellations with the R\&S FSW43 analyzer. As previously elucidated in Fig.\;\ref{fig:drifts}(b)-(c), the central frequencies of the OMCC harmonics display some variability, with this frequency offset escalating with the harmonic order. For simplification, the performance of our 3GPP 5G NR receiver is evaluated using a 100-MHz BW 5G NR signal centered at $f_{0}=1\;$GHz, with the demodulator's center frequency adjusted to $f_{0}+f$\textsubscript{offset}. Figure\;\ref{fig:Constellations}(a) confirms that frequency offsets below 8\;kHz exert negligible impact on the received constellation, allowing clear identification of two orange points corresponding to binary phase shift keying (BPSK) pilots and four blue points of the PDSCH data channel. It should be noted that for offsets higher than 100\;Hz the analyzer should be configured to demodulate QPSK data signals, as for offsets between 100\;Hz and 8\;kHz, it tends to miscalculate the quality of the signal by considering some constellation points as higher-order modulation (e.g. 256QAM). For offsets higher than 8\;kHz, a phase mismatch is intensified in the R\&S FSW43 (which is more sensitive than 3GPP 5G NR recommendation\cite{3GPP}), showing an outer ring similar to an amplitude phase shift keying (APSK) signal. This phenomenon was also observed in Fig.\;\ref{fig:upconversion}(b) when upconverting using the second harmonic of the OM frequency comb (with $k=2$). 

Fig.\;\ref{fig:Constellations}(b) shows the received EVM and constellations of two upconverted signals utilizing different harmonics while maintaining a constant SNR$=25\;$dB. It can be observed that the received signal upconverted using the second harmonic has a higher EVM compared with using the first harmonic, although taken with similar SNR. As represented in Fig.\;\ref{fig:drifts}, higher-order harmonics possess elevated phase noise and greater frequency drift. Also, higher-order PDSCH modulations were evaluated experimentally with quadrature amplitude modulation (QAM). As it can be observed in Fig.\;\ref{fig:Constellations}(c), a successful upconversion of 16QAM 5G NR signals is achieved with the first harmonic of the OMCC, ensuring an EVM better than the 3GPP recommendation of 12.5\% for 16QAM modulation\cite{3GPP}. Remarkably, a slightly higher SNR is required compared with QPSK to achieve a similar EVM. As part of ongoing work, the implementation of a closed loop, wherein the output of the OMCC feeds back into the input, could be explored to reduce the frequency drift and enhance the system stability. Refining the oscillator has the additional benefit of increasing the peak power of the lasing OMCC mode, potentially increasing the power of each harmonic and the resulting SNR, also enabling higher-order modulations. 

\subsection{Impact of signal-to-noise ratio}
\begin{figure}[b]
  \centering
\includegraphics[width=8cm]{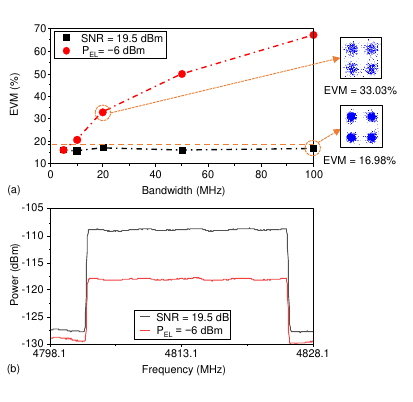}
     \vspace{-0.7cm}
\caption{\fontsize{8}{10}\selectfont\centering (a)\;Received EVM of a 5G NR signal upconverted to $4.8\;$GHz with constant electrical power of the signal generator $P$\textsubscript{EL}$=-6$\;dBm and with constant SNR$=19.5$\;dB. (b)\;Electrical spectrum corresponding to 20-MHz BW 5G upconverted signals for SNR=19.5\;dB vs. $P$\textsubscript{EL}$=-6$\;dBm.}
\label{fig:SNRPower}
\end{figure}

Next, we studied the influence of the received SNR on the resulting EVM of the frequency-converted 5G signals. For simplicity, we evaluated the performance of the upconverted signals using the first harmonic of the OM frequency comb ($k=1$), which appears at center frequency $f_{m}+f_{0}=4.84$\;GHz. In Fig.\;\ref{fig:SNRPower}(a), two curves depicting BW versus EVM are presented: one curve maintains constant electrical power (set at P\textsubscript{EL}$=-6$\;dBm), while the other represents a scenario with constant SNR=19.5\;dB. Notably, the curve of constant SNR exhibits a marginal variation in EVM, ranging from 16.28\% for 5-MHz BW signals to 16.98\% with 100-MHz BW. In contrast, the curve with constant generated power displays a more pronounced variation, escalating from 16.28\% at 5\;MHz to 67.33\% at 100\;MHz, only satisfying the 3GPP 5G PDSCH recommendation for 5-MHz BW, which corresponds to an SNR of 19.5\;dB. This observation underscores that the deterioration in EVM is correlated with a reduction in SNR for the same electrical power over different BWs. Further clarification is provided in Fig.\;\ref{fig:SNRPower}(b), where the electrical spectra of the upconverted signals evaluated in Fig.\;\ref{fig:SNRPower}(a) are presented. In this case, we compare the spectrum of upconverted 20-MHz BW 5G NR signals when setting the SNR to 19.5\;dBm and when configuring the generator with constant power at P\textsubscript{EL}$=-6$\;dBm. The difference in SNR between both configurations is evident.  

\begin{figure}[!t]
  \centering
\includegraphics[width=8.5cm]{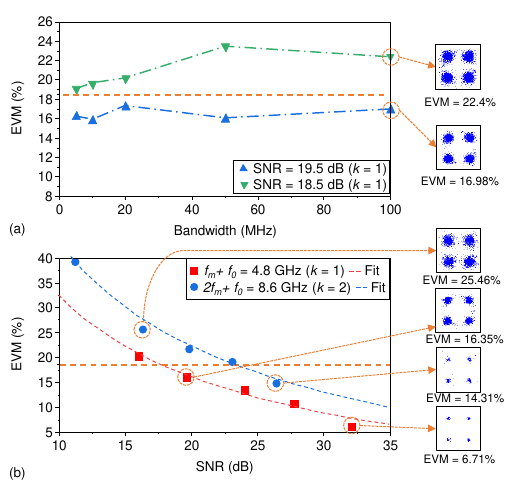}
     \vspace{-0.5cm}
\caption{\fontsize{8}{10}\selectfont\centering (a)\;Received EVM of the upconverted 5G NR signal with the first harmonic ($k=1$) measured at $f_{m}+f_{0}=4.8$\;GHz for different SNR and BWs. (b)\;Received mean EVM calculated from 3 measurements taken in a 10-minute interval with identical BWs for the upconverted 5G NR signal using the first and second harmonics for different SNRs. \mbox{3GPP EVM} recommendation included as a horizontal dashed line. Constellation examples included as insets marked with circles in the graphs.}
\label{fig:EVMSNR}
\end{figure}

In Fig.\;\ref{fig:EVMSNR}(a), the EVM evaluation of frequency upconversion using the first harmonic is evaluated with constant SNR. The experimental results confirm that, for SNR values above 19.5\;dB, all the BWs meet the EVM criteria. In Fig.\;\ref{fig:EVMSNR}(b), two sets of data are presented to further evaluate the relationship between SNR and EVM, each one corresponding to the received upconverted 5G NR signal using the first and second harmonics provided by the OMCC. Within each dataset, each symbol on the graph represents the mean values of EVM and SNR calculated from three measurements taken in a 10-minute interval with identical BWs. The EVM behaviour is consistent with \cite{EVMSNRa,EVMSNRb}: EVM\;$\approx$ SNR$^{-1/2}$. Two fit lines have been introduced in Fig.\;\ref{fig:EVMSNR}(b) with the approximation of $y=a\cdot x^{-1/2}+b$ with a coefficient of determination $(R^2)$ higher than 0.96 for the first harmonic ($k=1$) and higher than 0.99 for the second harmonic ($k=2$). A discernible difference in EVM can be observed by comparing the signals upconverted with the first and with the second harmonic. This arises from the frequency offset and phase noise associated with the second harmonic, as depicted in Fig.\;\ref{fig:drifts}(b) and (c). Consequently, the signals upconverted with the second harmonic show a higher EVM than those upconverted with the first harmonic, even for similar received SNR values. Within the SNR range from 15\;dB to 25\;dB, this difference in EVM is approximately 6\%. This degradation implies that, whilst applications using the first harmonic require a received SNR of 19.5\;dB, using the second harmonic to achieve higher frequency bands requires SNR values above 23\;dB. 

\section{Conclusions}
This paper proposes and demonstrates experimentally the all-optical frequency conversion of 3GPP 5G NR signals using an on-chip OMCC that enables the up- and down-frequency conversion of microwave signals in the GHz range using the generated OM frequency comb. OMCCs main advantages includes the reduced footprint (ideal for massive integration on silicon chips) and minimal driving power (ideally below 1 mW), making it well-suited for low-power applications. These features significantly bolster the practicality and cost-effectiveness of our proposed all-optical frequency conversion method. The first and second harmonics generated by the OM frequency comb in the lasing regime mode were evaluated in detail in terms of amplitude span, frequency offset, and phase noise. A maximum frequency deviation of 12\;kHz for the first harmonic and 28\;kHz for the second harmonic have been measured over 10\;s (short-term evaluation). As expected, the stability of the mechanical resonance has a higher impact on higher-order harmonics of the generated OM frequency comb. Thus, the frequency offset of the second harmonic was measured to be approximately twice that of the first harmonic. The stability of the OMCC was measured for a 20-min interval (long-term evaluation), obtaining a frequency drift smaller than 250\;kHz for both harmonics. These frequency offset values meet the 3GPP specification for 5G NR signals with the maximum BW of the standard of 100\;MHz. The given OMCC results in low phase noise values of -97.1\;dBc/Hz and -91.95\;dBc/Hz at 100\;kHz for the first and second harmonics, respectively. This phase noise value is smaller than other Silicon photonic integrated solutions \cite{LowPhaseNoise}.\\
\indent Frequency upconversion from the 5G low-band at 1\;GHz to the 5G mid-band at 4.8\;GHz using the first harmonic is demonstrated with signals up to 100-MHz BW, with received EVM smaller than 9.84\%. Frequency upconversion using the second harmonic to the 5G extended mid-band and next-generation 6G at 8.6\;GHz is also achievable but with smaller BW, due to the reduced SNR using the second harmonic. Successful upconversion with the second harmonic can be achieved with up to 10-MHz BW 5G NR signals with EVM smaller than 17.68\%. The downconversion process from the same frequencies (from 8.6\;GHz back down to 4.8\;GHz) is also demonstrated with up to 100-MHz BW 5G NR signals with EVM smaller than 15.87\%, confirming the all-optical multi-frequency conversion using the OM frequency comb. \\
\indent Comparing the performance using the first and second harmonics, an EVM degradation of 6\% is observed for higher-order harmonics due to the frequency offset and phase noise difference between the harmonics generated by the OMCC. When using the first harmonic an SNR of 19.5\;dB is recommended, while using the second harmonic requires a SNR above 23\;dB. The reduction of SNR for higher-order harmonics may hinder the application of our technique at frequencies above 8 GHz (roughly, the second harmonic of our cavity). A possible solution would be the use of more complex, two-dimensional OMCCs supporting mechanical resonances at higher frequencies \cite{OMCHighFrequency1,OMCHighFrequency2}. In principle, such cavities should show similar phase noise levels, though no experimental results have been yet provided.  As ongoing work, a closed loop in which the output of the OMCC feeds back into the input could be implemented to reduce the frequency offset and improve the stability of higher-order harmonics.

\vfill

\end{document}